\documentclass[final,5p,times,twocolumn,authoryear]{elsarticle}

\usepackage{amssymb}
\usepackage{amsmath}

\usepackage{hyperref}

\journal{Journal of High Energy Astrophysics}

\begin{document}

\begin{frontmatter}



\title{Cooling of neutron stars in soft X-ray transients with realistic crust composition}

\author[Ioffe]{A.Y. Potekhin} 
\author[Ioffe]{A.I. Chugunov} 
\author[ULB]{N.N. Shchechilin} 
\author[Ioffe]{M.E. Gusakov} 

\affiliation[Ioffe]{organization={Ioffe Institute},
            addressline={Politekhnicheskaya 26}, 
            city={Saint Petersburg},
            postcode={194021},
            country={Russia}}
\affiliation[ULB]{organization={Institut d'Astronomie et
d'Astrophysique, ULB},
            addressline={CP226}, 
            city={Brussels},
            postcode={1050},
            country={Belgium}}

\begin{abstract}
Thermal radiation of neutron stars in soft X-ray transients (SXTs) in a
quiescent state is believed to be powered by the heat deposited in the
stellar crust due to nuclear reactions during accretion. Confronting
observations of this radiation with simulations helps to verify
theoretical models of the dense matter in  neutron stars. We simulate
the thermal evolution of the SXTs with theoretical models of the
equation of state and composition of the accreted crust.
The new family of such models were recently developed
within a thermodynamically consistent approach by
modeling the nuclear evolution of an accreted matter as it sinks toward
the stellar center, starting from representative thermonuclear ash
compositions. The crust cooling curves computed with the traditional and
modern theory are compared with observations of SXTs MXB 1659$-$29 and
IGR J17480$-$2446. We show that the new and traditional models of the
accreted neutron star crusts are similar in their capability to explain
the thermal evolution of neutron stars in SXTs. Both kinds of models
require  inclusion of additional ingredients not supplied by the current
theory, such as the shallow heating and variation of thermal
conductivity, to fit observations.
\end{abstract}


\end{frontmatter}


\section{Introduction}
\label{sec:intro}

Many neutron stars (NSs) reside in binary systems and accrete matter
from the companion. Some of them, called soft X-ray transients (SXTs),
accrete intermittently, so that accretion episodes (outbursts) alternate
with periods of quiescence, when accretion terminates or is strongly
suppressed. The outbursts are revealed by intense X-ray radiation due to
the release of the gravitational energy of matter falling from the
accretion disk on a NS. In the quiescence, the X-ray luminosity decreases by
orders of magnitude and may be determined by thermal emission from the
NS surface (see, e.g., \citealt{WijnandsDP17} for a review).

The accretion during an outburst leads to nuclear reactions in the crust
accompanied by heat release. Thermonuclear reactions in the outermost NS
envelopes transform accreted protons or alpha particles into a
complex mixture of heavy elements, called thermonuclear ashes. Most of
the energy produced in these reactions is immediately irradiated in
X-rays and neutrinos. According to  the deep crustal heating scenario
\citep{Sato79,HaenselZdunik90}, heat is also produced as the crust
matter is pushed inside the star under the weight of newly accreted
material. A part of the accumulated heat propagates from the NS crust to the
core and the other part leaks through the surface. The so-called
quasi-persistent SXTs have long outbursts (lasting months to years),
which appreciably warm up the crust. Thermal relaxation of the
overheated crust manifests itself by the decline of the observed thermal
X-ray luminosity (e.g., \citealt{WijnandsDP17} and references therein). 

An analysis of observations of the post-outburst cooling allows one to
constrain the thermal conductivity and heat capacity of the crust (e.g.,
\citealt{Rutledge_02,Shternin_07,PageReddy13}). Such an analysis is
complicated. As a rule, interpretation of the observed crust cooling
within the deep crustal heating scenario requires involving the
so-called ``shallow heating'' by some additional energy sources at
relatively  low  densities \citep{BrownCumming09}. Besides, some SXTs
may experience weak and variable residual accretion during quiescence
\citep[e.g.,][]{TurlioneAP15}.

A traditional approach to  calculation of the equation of state (EoS)
and the heat release in the crust
\citep{HaenselZdunik90,HaenselZdunik03,HaenselZdunik08,Fantina_18} is
based on the assumption that free neutrons move together with the nuclei
during accretion.  Recently, the theoretical models of the accreted
crust  have been revised by \citet[][hereafter
GC]{GusakovChugunov20,GusakovChugunov21}, who found that diffusion of
free neutrons results in their redistribution over the inner crust,
conforming to the  hydrostatic and diffusion equilibrium condition (the
nHD condition).  \citet{GusakovChugunov21} derived an universal formula
for the total heating efficiency in the fully accreted inner crust in
the nHD equilibrium. \citet[][hereafter
SGC]{ShchechilinGC21,ShchechilinGC22,ShchechilinGC23} performed a more
detailed study of the composition, EoS, and heat release in the accreted
crust with allowance for nuclear transformations during the accretion
under the nHD equilibrium condition.

In a previous paper \citep[][hereafter Paper~I]{PotekhinGC23} we
performed numerical  simulations of the thermal evolution of the SXTs
in the nHD equilibrium and compared the
results with the traditional approach and with observations. For the
latter, we considered a collection of quasi-equilibrium thermal
luminosities of the SXTs in quiescence \citep{PotekhinCC19} and also the NS
crust cooling in SXT MXB 1659$-$29 \citep{Parikh_19}. We found that the observed quasi-stationary
thermal luminosities of the SXTs can be equally well fitted (within the
observational uncertainties) using the
traditional and thermodynamically consistent models, provided that the
shallow heat diffusion into the core is taken into account. The observed
crust cooling in MXB 1659$-$29 could also be fitted in the frames of both
models, but the choice of the model affects the derived parameters
responsible for the thermal conductivity in the crust and for the
shallow heating.

The simulations in Paper~I were based on the model of the composition
and heating in the outer crust developed by \citet{GusakovChugunov21,GusakovChugunov24}.
In this model, the nuclear transformations occur, starting from
$^{56}$Fe, at some discrete values of pressure in the outer crust, at
the interface between the outer 
and inner crusts, and near the 
interface between the inner crust and the core. In the present work, we
explore the thermal evolution of SXTs during the outbursts and
quiescence, using the most recent SGC calculations, which start with
realistic compositions of thermonuclear ashes, instead of $^{56}$Fe, and
are continued into upper layers of the inner crust. This update cannot
affect the basic conclusions of Paper~I concerning the quasi-stationary
quiescent states of the SXTs, which depend mainly on the total
time-averaged heat injected into the core during many consecutive
outbursts, because a total deep crustal heating of the star
is similar in the GC and SGC models.
However, the individual crust cooling curves are sensitive to the
positions and intensities of the discrete heating sources and to the
details of the crust composition, therefore they could be substantially
different for the different models.

In the present paper we perform numerical simulations of the thermal
evolution of SXTs with the SGC crust models and compare the crust
cooling curves with each other and with observations of SXTs MXB
1659$-$29 and IGR J17480$-$2446, whose luminosities in quiescent,
active, and post-outburst cooling states have been well monitored. For
these simulations we construct analytical fits to the EoS in the inner
crust and compose tables of the deep crustal heating intensities and
composition of the inner and outer crusts in the mean ion approximation.

The paper is organized as follows. In Section~\ref{sect:accrust} we
describe the accreted crust models used in this research. In
Section~\ref{sec:cool} we describe the simulations of the thermal
evolution of the SXTs MXB 1659$-$29 and IGR J17480$-$2446, compare these
simulations with observations, and discuss the results. We summarize the
conclusions in Section~\ref{sec:concl}. \ref{sec:fits} presents
analytical fits to the pressure and the fraction of free neutrons in the
inner crust, which are employed in the simulations.

\section{Accreted crust models}
\label{sect:accrust}

Composition of the deep accreted NS crust depends on the initial
composition in the shallow layers, which are pushed toward the stellar
center during accretion. The initial composition is formed by
thermonuclear burning of accreted elements in the NS envelope. The
burning proceeds in different regimes (stable or unstable) depending on
the composition of the accreted matter  and the accretion rate. The SGC
models of an accreted NS crust are based on three representative
thermonuclear ash compositions. A composition presented by
\citet{Cyburt_16}, based on the hydrodynamic code \textsc{kepler}
\citep{Woosley_04}, has a wide distribution of elements with the most
abundant nuclides being in the iron group. If there is a sufficiently
large amount of hydrogen at the moment of burst ignition, the extended
rapid proton capture process could lead to the formation of the
palladium group elements \citep{Schatz_01}. Besides, some accreting NSs
exhibit so-called superbursts, which originate from thermonuclear
explosion of carbon accumulated in the crust and result in a narrow
distribution of nuclides with a peak in the elements of the iron group
\citep{KeekHeger11}. In the SGC approach, the composition and EoS of the
accreted crust is found by tracing pathways of nuclear transformations
with increasing pressure, using a
 nuclear reaction network and starting
with one of these three thermonuclear ash compositions  in shallow
depths. The three respective models of the accreted crust are briefly
called Kepler (K), Extreme rp (RP), and Superburst (SB) models.

The GC and SGC accreted crust models depend on the pressure
$P_\mathrm{oi}$ at the interface between the outer and inner crust.
Because of  the diffusion of neutrons, $P_\mathrm{oi}$ may differ from
the neutron-drip pressure for the pristine crust. 
  The value of $P_\mathrm{oi}$ is currently
unknown, and it is used as a free parameter in constructing the accreted
crust models. \citet{GusakovChugunov20,GusakovChugunov24} argued that $P_\mathrm{oi}$
should lie between the minimum value $P_\mathrm{oi}^\mathrm{(0)}$, which
corresponds to the case where no heat is released in the deep layers of
the inner crust beyond the scope of the detailed modeling, and maximum
$P_\mathrm{oi}^\mathrm{(cat)}$, which roughly equals the neutron-drip
pressure in cold catalyzed matter. The pressure values, at which the
nuclear transformations occur, as well as the values of
$P_\mathrm{oi}^\mathrm{(0)}$, $P_\mathrm{oi}^\mathrm{(cat)}$ and the
amounts of heat released at each transformation, depend on the initial
nuclear composition and on the employed nuclear models.

The SGC calculations were performed in frames of a simplified reaction
network \citep{ShchechilinGC21} and employed  theoretical mass tables
for the finite-range droplet macroscopic model of nuclei FRDM12
\citep{Moeller_16}. The values of $P_\mathrm{oi}^\mathrm{(0)}$ for the
K, RP, and SB models are 
$7.26\times10^{29}$
dyn cm$^{-2}$, 
$7.75\times10^{29}$ dyn cm$^{-2}$, and 
$7.29\times10^{29}$ dyn
cm$^{-2}$, respectively, and $P_\mathrm{oi}^\mathrm{(cat)}$ for the
FRDM12 model equals $7.86\times10^{29}$ dyn cm$^{-2}$. The SGC
calculations provide sets of K, RP, and SB crust models for
$P_\mathrm{oi}$ values in the range $6.4\times10^{29}\mbox{~dyn cm}^{-2}
\leq P_\mathrm{oi} \leq 10^{30}\mbox{~dyn cm}^{-2}$, which embraces the
interval $[P_\mathrm{oi}^\mathrm{(0)},P_\mathrm{oi}^\mathrm{(cat)}]$.
For each $P_\mathrm{oi}$ value, there is a distribution of nuclei,
depending on 
the outer crust composition. The latter 
has been computed and tabulated for the
layers with $P$  from  $6\times10^{26}$ dyn cm$^{-2}$ at
$\rho\approx(1.3-1.4)\times10^9$ g cm$^{-3}$ in the outer crust to 
$P\approx 2\times10^{30}$ dyn cm$^{-2}$ at $\rho=\rho_\mathrm{max}\equiv
2\times10^{12}$ g cm$^{-3}$ in the inner crust. At certain pressure
values, the nuclear transformations are accompanied by heat release. 

In the present work we employ the mean ion approximation: we calculate
the mean charge and mass numbers, $\langle Z\rangle$ and $\langle
A\rangle$, from the detailed tables obtained by SGC and plug them into
the analytical formulae for thermodynamic functions, thermal
conductivities, and neutrino emissivities \citep[][and references
therein]{PotekhinPP15}. Here and hereafter, the angle brackets denote an
average with  weights proportional to number fractions of the nuclei. In
addition to $\langle Z\rangle$ and $\langle A\rangle$, the thermal
conductivities depend on the ``impurity parameter''
$Q_\mathrm{imp}\equiv \langle (Z - \langle Z\rangle)^2 \rangle$, which
is also obtained from the detailed SGC tables.
At shallow depths above the first nuclear transformation ($P
\lesssim 6\times10^{26}$ dyn cm$^{-2}$, $\rho\lesssim1.4\times10^9$ g
cm$^{-3}$),
we assume the same composition 
as 
in the initial mixture of the specific model.
In particular, the initial composition affects the temperature profile in the
heat blanketing envelope, making the relation between the internal
temperature in the core and the
effective surface temperature $T_\mathrm{eff}$ in quiescence slightly
different for the different ash models.
The fraction of free
neutrons and the EoS in the inner crust are given by the analytical
fitting formulas in \ref{sec:fits}. These fits provide a good accuracy
in the entire range of the densities covered by the SGC data for the
inner crust, $\rho_\mathrm{oi+}<\rho<\rho_\mathrm{max}$, where
$\rho_\mathrm{oi+}$ is the density at the outer edge of the inner
crust. 

\begin{figure}[t!]
\centering
\includegraphics[width=\columnwidth]{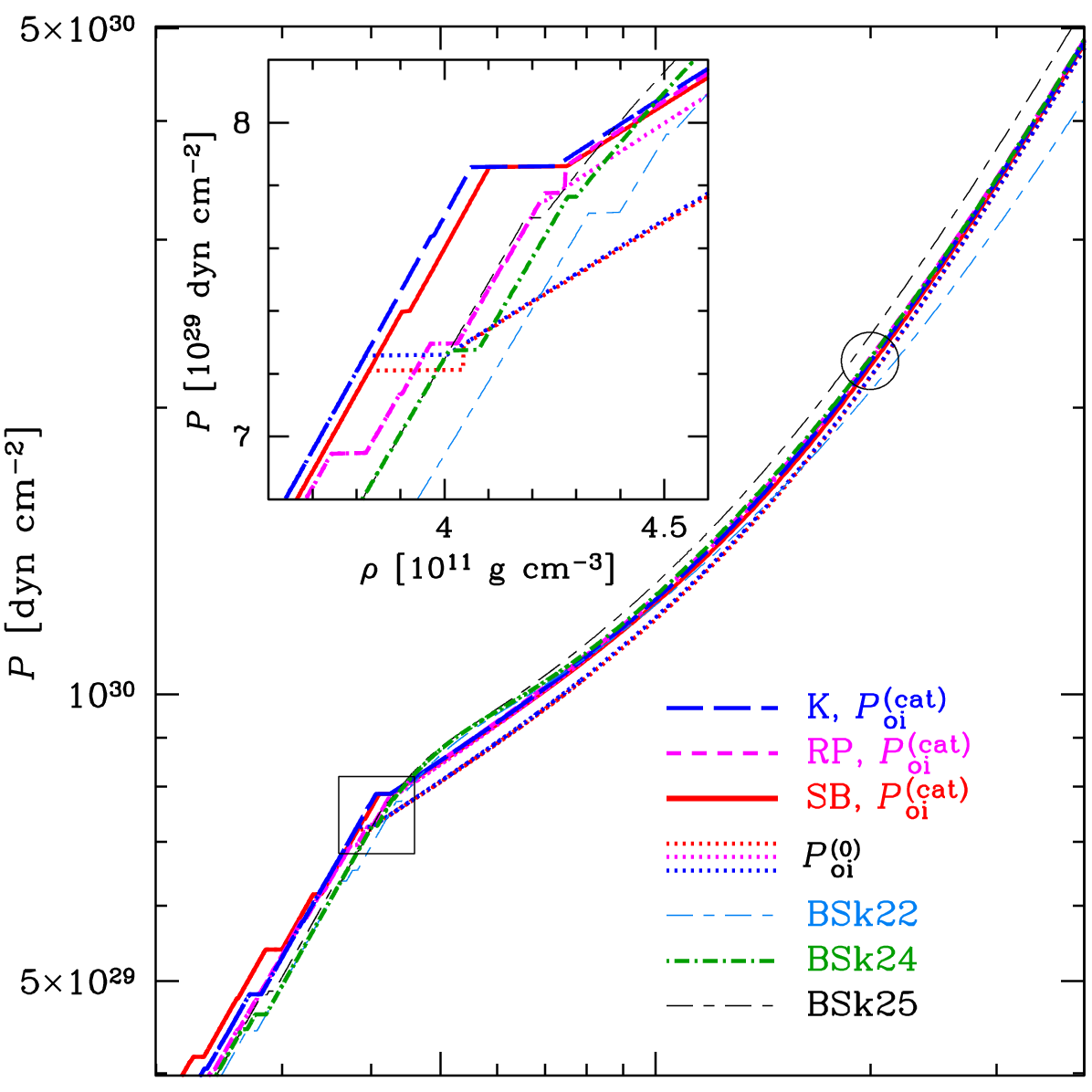}
\includegraphics[width=\columnwidth]{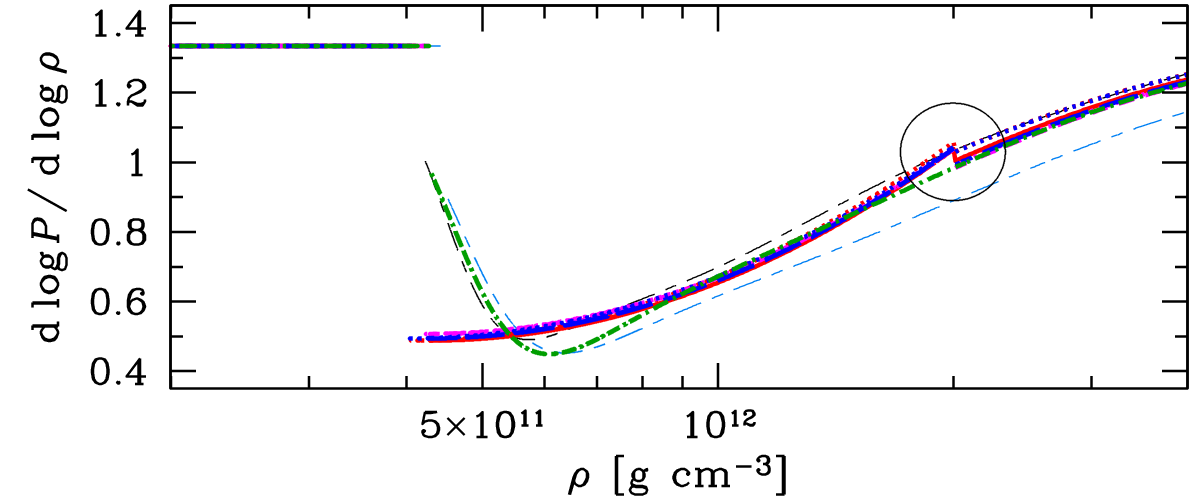}
\caption{EoS models. \emph{Upper panel:} Pressure $P$ as
function of mass density $\rho$ is plotted according to the tabular
composition in the outer crust and analytic fits in the inner crust,
extended beyond the computed data range
as explained in the text. Red solid, magenta short-dashed,
and blue long-dashed lines show the EoS with the Kepler (K), Extreme rp
(RP),and Superburst (SB) models for the accreted crust, respectively,
with $P_\mathrm{oi}=P_\mathrm{oi}^\mathrm{(cat)}$. The dotted curves of
the respective colors correspond to the same models with
$P_\mathrm{oi}=P_\mathrm{oi}^\mathrm{(0)}$. For comparison, the EoSs of
cold catalyzed NS matter are shown: BSk24 (dot-dashed line), BSk22 and
BSk25 (lower and upper thin long-and-short-dashed lines). The inset
shows a zoom to the vicinity of the outer-inner crust interface,
selected by the rectangle. \emph{Lower panel:} Logarithmic derivative
of pressure with respect to the density for the
same EoS models. The gaps correspond to the outer-inner
crust interface. In both panels, the circle highlights the region of
maximum $P$ and $\rho$ values for the computed SGC data, where the
fit to the data is matched by a shifted BSk24 EoS (see text).
}
\label{fig:EoS}
\end{figure}

\begin{figure}[t]
\centering
\includegraphics[width=\columnwidth]{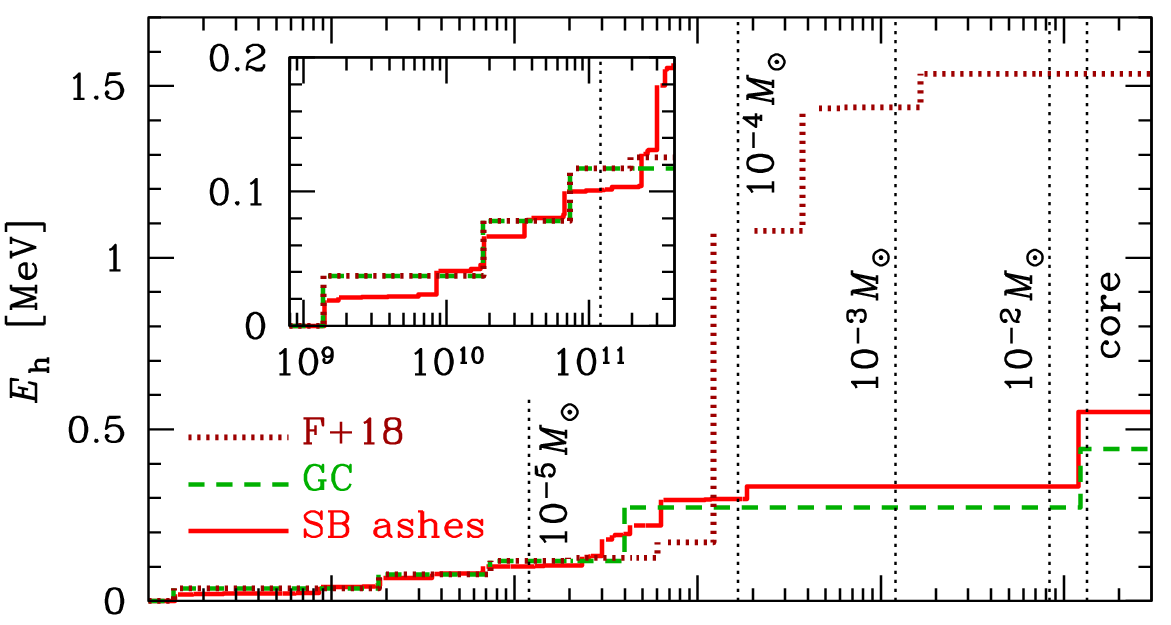}
\includegraphics[width=\columnwidth]{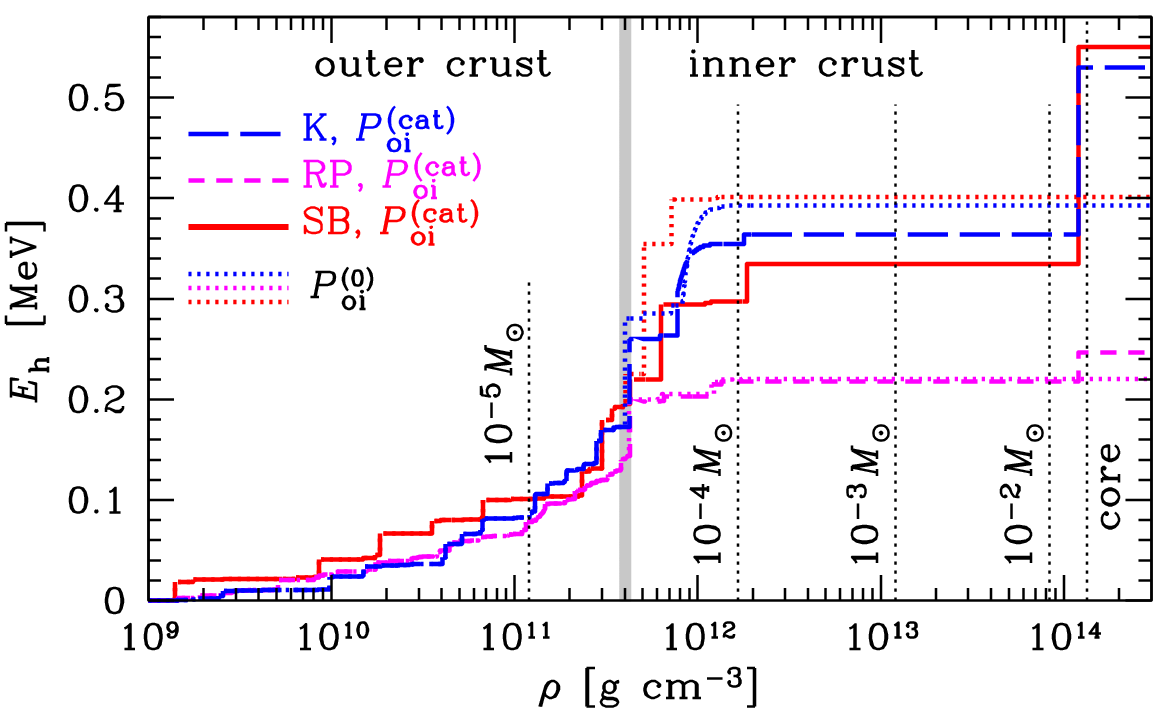}
\caption{\emph{Upper panel}: Total heat $E_\mathrm{h}$, generated per accreted baryon 
in a layer from the star surface to a given mass
density $\rho$, according to the model of \citet{Fantina_18} (F+18,
dotted line), the model of \citet{GusakovChugunov21} with
$P_\mathrm{oi}=P_\mathrm{oi}^\mathrm{(cat)}$ (GC, dashed line),
and the model of \citet{ShchechilinGC23} with
$P_\mathrm{oi}=P_\mathrm{oi}^\mathrm{(cat)}$ for the SB ashes (solid line). The gaps on the lines (best
visible in the F+18 case) are due to the density jumps at the interfaces
between neighboring layers containing different nuclei. The vertical dotted lines mark the $\rho$ values
corresponding to the mass of overlying material, from
$10^{-5}\,M_\odot$ to $10^{-2}\,M_\odot$, labeled near these lines, for
a NS with mass $M=1.6\,M_\odot$ and radius
$R=12.6$~km, which corresponds to BSK24 EoS in the core. 
\emph{Lower panel}: The same for six SGC models: K, RP, and SB with
$P_\mathrm{oi}=P_\mathrm{oi}^\mathrm{(cat)}$ (long-dashed blue, 
short-dashed magenta, and solid red lines, respectively) and with
$P_\mathrm{oi}=P_\mathrm{oi}^\mathrm{(0)}$ (dotted lines of the
respective colors). The vertical gray strip shows the range of the
density at outer-inner crust boundary for the selected models,
$\rho_\mathrm{oi}\approx(3.8-4.3)\times10^{11}$ g cm$^{-3}$.
}
\label{fig:heat1}
\end{figure}

In the deeper crust layers, at $\rho>\rho_\mathrm{max}$, the numerical
SGC data are absent. It turns out, however, that the different EoSs,
corresponding to different values of $P_\mathrm{oi}$, become close to
the EoS model BSk24 \citep{Pearson_18} with increasing $\rho$ towards
$\rho_\mathrm{max}$. Therefore, for the sake of the NS modeling, we
extend the fitted EoS at $\rho > \rho_\mathrm{max}$ using the BSk24 EoS,
modified by a constant shift of pressure.
This shift is applied so as to get a continuous
$P(\rho)$ at the matching point:
$P(\rho)|_{\rho>\rho_\mathrm{max}}=P_\mathrm{BSk24}(\rho) +
P_\mathrm{fit}(\rho_\mathrm{max})-P_\mathrm{BSk24}(\rho_\mathrm{max})$,
where $P_\mathrm{fit}$ is the fit (\ref{fitP}) for the SGC EoS and
$P_\mathrm{BSk24}$ is the fit of \citet{Pearson_18} for the BSk24 EoS.
Examples of the resulting EoS are shown in the upper panel of
Fig.~\ref{fig:EoS}. For comparison, three unified EoSs of the cold
catalyzed crust are also shown: BSk22, BSk24, and BSk25
\citep{Pearson_18}, which differ by the assumed nuclear symmetry energy
at the nuclear saturation density
(32 MeV, 30 MeV, and 29 MeV, respectively). The lower panel demonstrates
the logarithmic derivative
$\chi_\rho\equiv\mathrm{d}\log\,P/\mathrm{d}\log\,\rho$.  The matching
point manifests itself by a break of $\chi_\rho$ (encircled). The break
is rather small, and the extended EoS fits to the accreted crust models
become indistinguishable from the BSK24 EoS at large densities
$\rho\gg\rho_\mathrm{max}$.
The same EoS was employed to describe the pure neutron matter outside clusters in SGC models.

In the upper panel of Fig.~\ref{fig:heat1} we compare predictions of one
traditional model and two nHD-consistent models GC and SGC for the heat
generated per accreted baryon, $E_\mathrm{h}$, integrated from the
surface to a given density in the crust, as a function of mass density.
As an example of a traditional model (without the nHD equilibrium), we
have chosen the relatively recent and advanced model by
\citet{Fantina_18}, specifically its version consistent with the BSk24
energy density functional (for short, the F+18 model). As an example of
the GC model, we have chosen the version that is consistent with the
BSk24 energy density functional and assumes
$P_\mathrm{oi}=P_\mathrm{oi}^\mathrm{(cat)}=7.731\times10^{29}$ dyn
cm$^{-2}$. The SGC models are presented by the SB ashes version with
$P_\mathrm{oi}=P_\mathrm{oi}^\mathrm{(cat)}$. 

The heating function $E_\mathrm{h}(\rho)$ is qualitatively similar for
the GC and SGC models, although the SGC model reveals more details:
there are more heating layers in the outer crust (at
$\rho<\rho_\mathrm{oi}\sim4\times10^{11}$ g cm$^{-3}$)  and, in
addition, some heating sources in the upper part of the inner crust (at
$\rho_\mathrm{oi}<\rho<\rho_\mathrm{max}$). 
The lower panel of Fig.\ref{fig:heat1} presents a comparison of the functions
$E_\mathrm{h}(\rho)$ for the K, RP and SB versions of the SGC model with
$P_\mathrm{oi}=P_\mathrm{oi}^\mathrm{(cat)}$ and
$P_\mathrm{oi}=P_\mathrm{oi}^\mathrm{(0)}$.

A total deep crustal heating of the star, $E_\mathrm{h}^\mathrm{(tot)}$
is given by $E_\mathrm{h}$ at $\rho\geq\rho_\mathrm{cc}$, where
$\rho_\mathrm{cc}\sim(1-2)\times10^{14}$ g cm$^{-3}$ is the density of
the crust-core interface. It is given by the line segments near the
right axes in Fig.~\ref{fig:heat1}. We see that the GC and SGC models
predict $E_\mathrm{h}^\mathrm{(tot)}$ in the range of $\sim0.2$--0.6 MeV
per an accreted baryon. It is much  weaker than in the traditional
models (such as F+18), which predict $E_\mathrm{h}^\mathrm{(tot)} \sim
(1.5-2)$~MeV. In Paper~I, we have already examined the influence of such
a difference in $E_\mathrm{h}^\mathrm{(tot)}$ on the quasi-stationary
thermal states of the SXTs in quiescence. The effect of this difference
turned out to be not very large in view of the observational and
theoretical uncertainties. In particular, the effect is smeared by the
shallow heating, which contributes in warming up the NS core due to the
heat transport through the crust. Clearly, the effect of the difference
between the GC and SGC models on the quasi-stationary SXT states should
be still smaller than the effect of their difference from the
traditional models. For this reason, in the present work we do not study
the quasi-stationary thermal states of the SXTs, but focus on the
dynamics of post-outburst crustal cooling.

\section{Cooling of SXT crusts}
\label{sec:cool}

To explore the thermal evolution of the SXT crusts with the SGC models,
we have chosen two SXTs, MXB 1659$-$29 and IGR J17480$-$2446, whose
crust cooling observations are sufficiently complete to reduce a
number of uncertainties: their crustal cooling has been well monitored,
the start and end dates of the outbursts are known, quasi-equilibrium
effective temperatures are measured with a sufficient accuracy, and average
accretion rates during outbursts have been estimated. We perform
numerical simulations of the crust heating during an outburst and the
subsequent cooling during quiescence, using the code described in
\citet{PotekhinChabrier18} with essentially the same physics input as in
Paper~I, except for the new crust models.

\begin{figure*}[t]
\centering
\includegraphics[width=1.3\columnwidth]{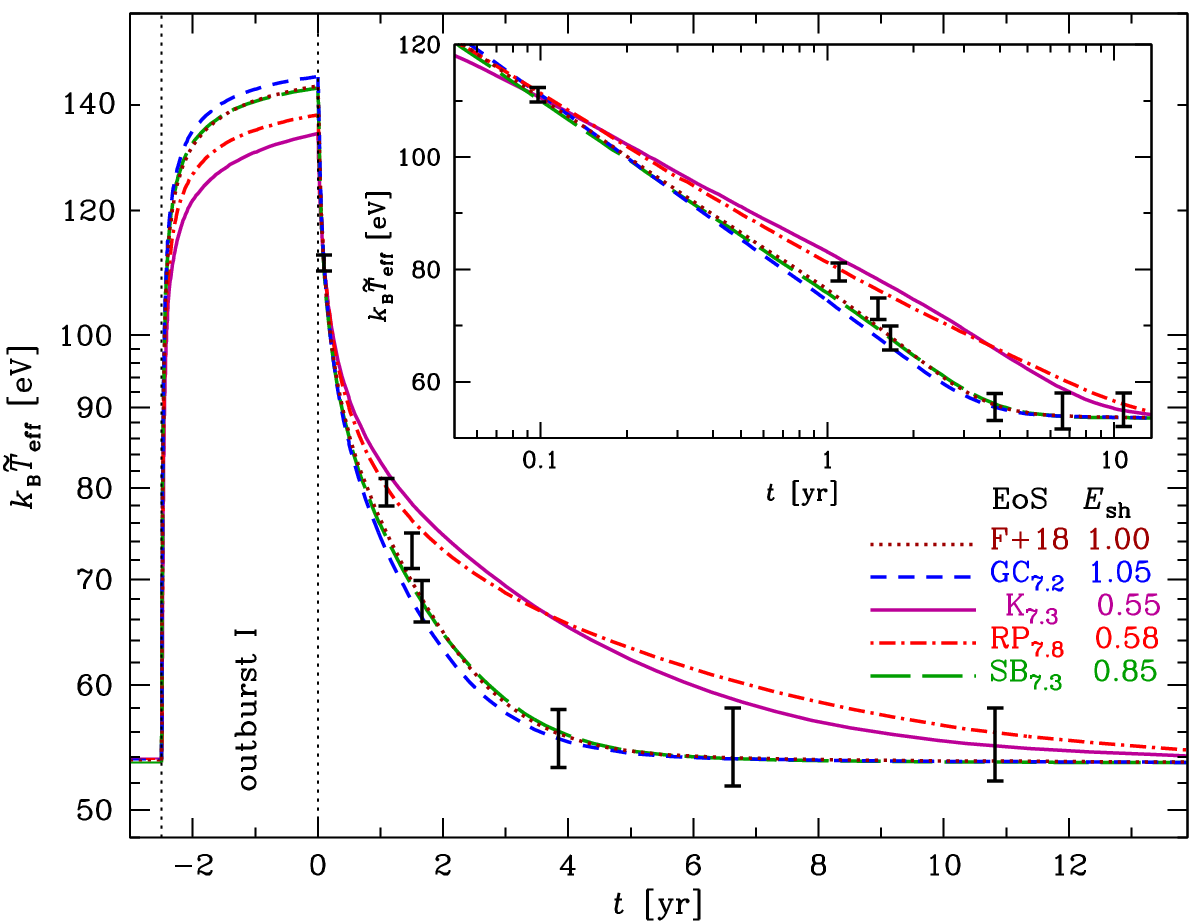}
\includegraphics[width=.629\columnwidth]{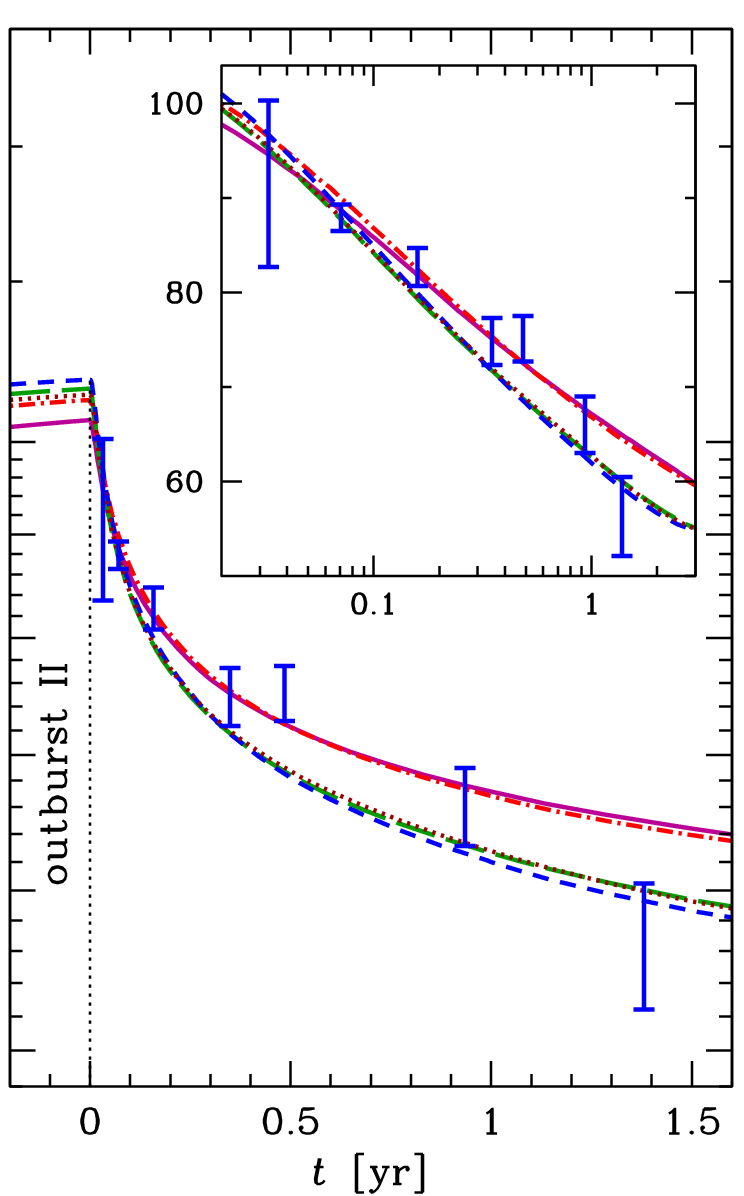}
\caption{Simulated light curves for the outbursts I (\emph{left panel})
and II  (\emph{right panel}) of MXB 1659$-$29 versus observations. The
redshifted effective temperature in energy units
$k_\mathrm{B}\tilde{T}_\mathrm{eff}$ is shown as function of time $t$,
counted from the end of an outburst. The dotted line  is calculated
assuming the F+18 model, and the other four lines represent different
nHD-consistent models   with $P_\mathrm{oi}=P_\mathrm{oi}^\mathrm{(0)}$.
The short-dashed line represents the results for the GC model, while the
solid, dot-dashed, and long-dashed lines correspond to the  SGC models
K, RP, and SB. The subscripts at the  symbols GC, K, RP, and SB in the
legend indicate the approximate values of $P_\mathrm{oi29}$ for the
shown models. The vertical errorbars (black and blue for the crust
cooling after outbursts I and II respectively) show the spectral fitting
results from \citet{Parikh_19} at the 90\% confidence, except for the
bar at $t = 10.8$ yr in the left panel, which represents one of the
different estimates by \citet[][see text]{Cackett_13}. The vertical
dotted lines separate the crust heating during an outburst from the
passive cooling during the quiescence. The insets show the same
dependences  and data at another scale for the crust cooling stage only.
}
\label{fig:MXBa}
\end{figure*}

\subsection{MXB 1659$-$29}
\label{sec:MXB}

The SXT MXB 1659$-$29 (MXB 1658$-$298, MAXI J1702$-$301) is a good touchstone for
testing crust cooling models, because it has been observed for a long
time, revealed three outbursts, and has a well documented record of the
crust cooling after the last two of them (see \citealt{Parikh_19} and
references therein). As in Paper~I, the accretion episodes observed in
1976--1979, 1999--2001, and 2015--2017 will be named \emph{outburst 0}, 
\emph{outburst I}, and \emph{outburst II}, respectively. The crust
cooling after the end of outburst I was simulated in a number of works
\citep{BrownCumming09,Cackett_13,Deibel_17,Parikh_19,PotekhinChabrier21,Mendes_22,PotekhinGC23,AllardChamel24a,AllardChamel24b}.
A consistent modeling of the short-term evolution of
MXB 1659$-$29 during and after both outbursts I and II was performed in
\citet{Parikh_19,PotekhinChabrier21}, and in Paper~I.
Here the term \emph{consistent modeling}
\citep[following][]{Parikh_19} means that all model parameters are kept
fixed and the same during and after different outbursts
(\citealt{AllardChamel24b} modeled the two bursts separately, without
requiring the constancy of the parameters).

The simulation procedure is similar to that in
\citet{PotekhinChabrier21} and in Paper~I. Starting from a quasi-steady
thermal structure of the NS with a  sufficiently hot interior, we
simulate passive cooling until the observed quasi-equilibrium luminosity
in quiescence is reached. For the latter, we fix the effective
temperature $\tilde{T}_\mathrm{eff}$ at a level consistent (within
uncertainties) with  observations of the SXT in quiescence. Here, tilde
indicates that the quantity is corrected for gravitational redshift. The
stellar mass was fixed,  $M=1.6 M_\odot$, to agree with quiescent
luminosity of MXB 1659$-$29 for average accretion rate, see figure 2 in
Paper~I. For the BSk24 EoS in the core, it results in the
stellar radius $R\approx12.6$~km.

Let us note that the
last observation
before outburst II (in 2012, after $t=10.8$ yr of cooling) gave
contradictory results, reported by \citet{Cackett_13}: the same spectral
analysis as for the observations at $t=6.6$ yr \citep{Cackett_08}
suggested $\tilde{T}_\mathrm{eff}$ consistent with the preceding  result,
but the count rate dropped. The authors tried different spectral models
by allowing the hydrogen column density $N_\mathrm{H}$ to differ from
the preceding value or by including a power-law spectral component in
addition to the thermal one, and obtained estimates of
$k_\mathrm{B}\tilde{T}_\mathrm{eff}$  (where $k_\mathrm{B}$ is the Boltzmann
constant) ranging from $43\pm5$ eV to $56\pm2$ eV.  Here
we adopt the quasi-stationary level at $k_\mathrm{B}\tilde{T}_\mathrm{eff}=54$~eV,
compatible with the observations at $t=6.6$ yr \citep{Cackett_08}.

Then we simulate outburst 0, assuming it to be the same as outburst~I,
during 2.5 years, followed consecutively by 20 years in quiescence, then
by the outburst~I, by 14 years of quiescence, by 1.52 years of
outburst~II, and by 
present day quiescence.  The accretion rate during
outburst~I is adopted at the level
$\dot{M}_\mathrm{I}=4\times10^{-9}\,M_\odot$ yr$^{-1}$, in accordance with the
estimate $\dot{M}_\mathrm{I}=(4\pm2)\times10^{-9}\,M_\odot$ yr$^{-1}$ in
\citet{PotekhinChabrier21}. The large uncertainty of this estimate is
mainly related to the uncertainty in the distance to this SXT,
which can be $9\pm2$ kpc or $12\pm3$ kpc depending on the assumed
composition of thermonuclear fuel in its X-ray bursts
\citep{Galloway_08}.

The observed accretion rate during outburst~II
in known to vary strongly, but we
simplify the model by adopting the constant rate
$\dot{M}_\mathrm{II}=1.8\times10^{-9}\,M_\odot$ yr$^{-1}$, as it provides the
best fits of the simulated cooling curves to the observations of the
crust cooling after outburst~II and roughly agrees with Fig.~1 of
\citet{Parikh_19}, which suggests the average accretion rate ratio
$\langle\dot{M}_\mathrm{II}\rangle/\langle\dot{M}_\mathrm{I}\rangle\sim1/3-1/2$.
The crust composition and shallow heating parameters are kept the same
for all three outbursts and cooling periods for each simulation set.

In most of our previous simulations of the thermal
evolution of MXB 1659$-$29, we assumed the position of the shallow
heating coincident with the position of the shallowest of the deep
heating sources at $\rho\approx1.4\times10^9$ g cm$^{-3}$. Here we adopt
the same rule. In Paper~I we demonstrated that one can produce almost
identical light curves by decreasing the depth of the shallow heating
layer to $\rho\sim10^8$ g cm$^{-3}$ with simultaneous increase of the
shallow heating power by 30--40\%. We adjusted the shallow heating
energy per baryon, $E_\mathrm{sh}$, so as the light curve matched the
first observation after outburst~I (at $t\approx0.1$ yr).

\begin{figure}[t]
\centering
\includegraphics[width=\columnwidth]{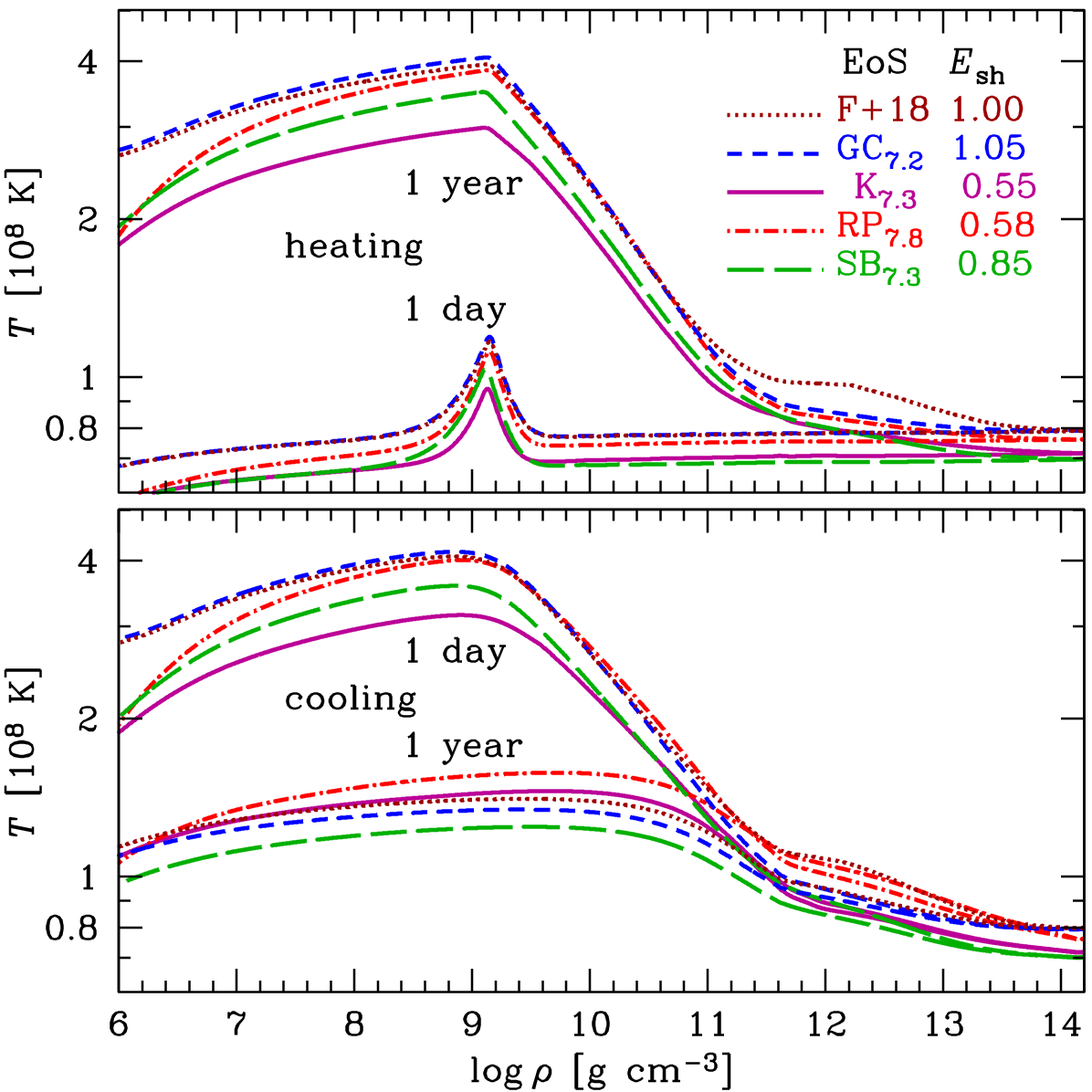}
\caption{Temperature profiles during the outburst~I (upper panel) and
subsequent cooling (lower panel) at 1 day and 1 year after the start of
the heating and the cooling, respectively. The same NS models as in
Fig.~\ref{fig:MXBa} are shown with the same line styles.
}
\label{fig:profiles}
\end{figure}

In Fig.~\ref{fig:MXBa} we compare the light curves  calculated using the
traditional model F+18, a GC model, and SGC models K, RP, and SB with
observations. For the GC and SGC models, we used the parameters
$P_\mathrm{oi}=P_\mathrm{oi}^\mathrm{(0)}$  (approximate values of
$P_\mathrm{oi29}\equiv P_\mathrm{oi}/10^{29}$ dyn cm$^{-2}$ are marked
by the subscripts in the legend).  We show the effective temperature
corresponding to the thermal flux from the outer crust to the surface
during and after the outbursts I and II, discarding the instantaneous
energy release near the NS surface due to the accretion during an
outburst. In each case, time is counted from the end of the outburst:
$t=0$ corresponds to MJD 52162 and 57809.7 for outbursts I and II,
respectively \citep{Parikh_19}. The vertical errorbars are the
observational estimates at the 90\% confidence level \citep{Parikh_19}.
For the observation at $t=10.8$ yr after outburst~I, we show the
estimate $k_\mathrm{B}\tilde{T}_\mathrm{eff}=55\pm3$~eV, obtained by
\citet{Cackett_13} using a spectral model with free $N_\mathrm{H}$ but
without a power-law component.

The models F+18 and SB produce cooling curves, compatible with
most observations, except for the ones at $t=1.1$ yr after outburst I
and at $t=0.5$ yr after outburst II, for which the errorbars lie
significantly higher than the curves. Although the F+18 and SB
curves are almost identical, they are produced using different shallow
heating, $E_\mathrm{sh}=1$ MeV and 0.85 MeV, respectively. The smaller
value in the SB case is needed to match the early cooling observation,
because of the smaller conductivity due to the non-zero impurity
parameter $Q_\mathrm{imp}$, corresponding to a mixture of different
nuclei. Indeed, it is known that an increase of $Q_\mathrm{imp}$
requires a decrease of $E_\mathrm{sh}$ to match the same early cooling
data (cf.~Paper~I).

\begin{figure}[t]
\centering
\includegraphics[width=\columnwidth]{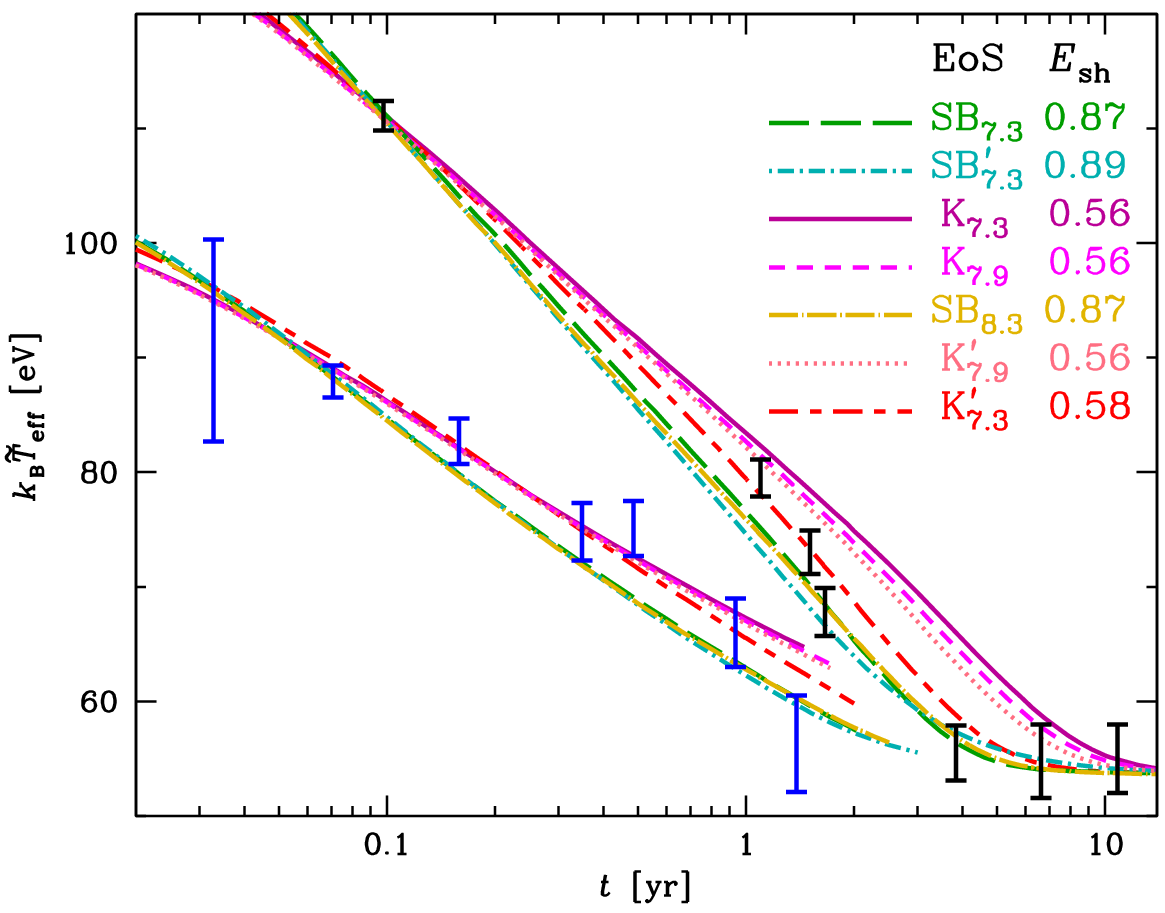}
\caption{Simulated light curves for the crust cooling after the
outbursts I (upper curves and black errorbars) and II (lower curves and
blue errorbars) of MXB 1659$-$29 versus observations. The data and
curves denoted SB$_{7.3}$ and K$_{7.3}$ are the same as in
Fig.~\ref{fig:MXBa}. The other curves are variations of the K and SB
models, denoted according to the legend: K$_{7.9}$ and SB$_{8.3}$ are
the models with higher $P_\mathrm{oi}$,  SB$_{7.3}'$ is characterized by
a suppressed neutron superfluidity, K$_{7.9}'$ is characterized by a
suppressed parameter $Q_\mathrm{imp}$ in deep layers of the crust, and
K$_{7.3}'$ is a model with a thin accreted crust (see text for details).
}
\label{fig:MXBb}
\end{figure}

The GC model with $Q_\mathrm{imp}=0$ requires a larger $E_\mathrm{sh}$
and produces a quicker crustal cooling because of the absence of heating
sources in the upper layers of inner crust. Conversely, the K and RP
models require a weaker shallow heating and produce a slower cooling
because of their much larger $Q_\mathrm{imp}$ (see Fig.~4 in
\citealt{ShchechilinGC23}). The latter two models appear to be
incompatible with observations of outburst I (left panel of
Fig.~\ref{fig:MXBa}).

Fig.~\ref{fig:profiles} provides  an insight into the thermal structure
of the NS models shown in Fig.~\ref{fig:MXBa}. We see that this
structure is dominated by the shallow heating. The slightly different
temperatures at high densities ensure the same quasi-equilibrium
$T_\mathrm{eff}$ for different models in quiescence, because of the
different composition of the heat blanketing envelopes, as mentioned in
Section~\ref{sect:accrust}.

In Fig.~\ref{fig:MXBb} we compare the post-outburst cooling for several
modifications of the K and SB models (the results for RP models are
similar to those for the K models). The curves marked SB$_{7.3}$ and
K$_{7.3}$ are the same as in Fig.~\ref{fig:MXBa}. The models SB$_{8.3}$
and K$_{7.9}$ are characterized by larger $P_\mathrm{oi}$. 
According to SGC,
the increase
of $P_\mathrm{oi}$ is accompanied by changes of the impurity parameter
in the inner crust:  $Q_\mathrm{imp}$ at $\rho=\rho_\mathrm{max}$ is
increased from 1.2 to 3.0  in the SB case and decreased from 5.4 to 3.8
in the K case. By default we keep $Q_\mathrm{imp}$ at
$\rho>\rho_\mathrm{max}$ equal to its value at 
$\rho=\rho_\mathrm{max}$. This may be an overestimation, because 
$Q_\mathrm{imp}$ demonstrates a decrease towards high densities. To test
a possible effect of this overestimation, we consider the model
K$_{7.9}'$, which differs from K$_{7.9}$ by setting $Q_\mathrm{imp}$ to
zero at $\rho>\rho_\mathrm{max}$. The model SB$_{7.3}'$ differs from
SB$_{7.3}$ by suppressing the singlet neutron superfluidity, which leads
to an increase of the specific heat in the inner crust. We see that
these modifications have minor effects on the crust cooling curves. In
all the cases the SB models, which have relatively small
$Q_\mathrm{imp}$, tend to underestimate the effective temperature at
intermediate crust cooling times, while the K models, which have
relatively large $Q_\mathrm{imp}$, tend to overestimate it.

Finally, we test the possibility that the reprocessed accreted matter
replaces the cold catalyzed matter only in the upper layers of the crust
to some density $\rho_\mathrm{ac} < \rho_\mathrm{cc}$
\citep[see][]{WijnandsDP13,Fantina_18,PotekhinCC19}. The model
K$_{7.3}'$ is constructed for the K composition with
$P_\mathrm{oi}=P_\mathrm{oi}^\mathrm{(0)}$, as in the model K$_{7.3}$,
but only at $\rho<\rho_\mathrm{ac}=10^{11}$ g cm$^{-3}$. At larger
densities, the BSk24 model is used. The restriction of the
accreted layer by the relatively small densities accelerates the
cooling, because thermal diffusion from shallow layers proceeds relatively
quickly. For this reason, the above-mentioned overestimation of the
effective temperature is reduced, so that the cooling curves for the
model with a partially accreted crust K$_{7.3}'$ are in better agreement
with observations than the K$_{7.3}$ curves for the fully accreted
crust.

\subsection{IGR J17480$-$2446}
\label{sec:IGR}

The SXT IGR J17480$-$2446 (Ter 5 X-2) is located in the globular cluster
Terzan 5. The distance to this cluster is known ($d=5.5\pm0.9$ kpc,
\citealp{Ortolani_07}), which reduces uncertainties in the analysis. 
This SXT revealed a relatively short (11 weeks, as estimated by
\citealt{DegenaarWijnands11}) outburst in 2010. It is the only outburst
confidently identified from this source.\footnote{As noted by
\citet{Ootes_19} (see also
\citealt{Heinke_ea24}),  some of the old outburst activity observed from
the direction of Terzan 5 may have been due to IGR J17480$-$2446 as
well, but the spatial resolutions of the earlier X-ray missions were not
sufficient to resolve sources in dense globular cluster cores. Here we
do not include this possibility into the analysis. The average accretion
rate during the outburst was estimated to be
$0.11\dot{M}_\mathrm{Edd}\approx3\times10^{-9}\,M_\odot$ yr$^{-1}$,
where $\dot{M}_\mathrm{Edd}$ is the Eddington accretion rate
\citep{DegenaarWijnands11,Ootes_19}. Thus the mean accretion rate over
$\sim30$ years of observations is $\sim2\times10^{-11}\,M_\odot$
yr$^{-1}$ \citep{PotekhinCC19}.}

The crust of the NS in IGR J17480$-$2446 had not
yet been relaxed by the time of its last published observation, but its
quasi-equilibrium effective temperature
$k_\mathrm{B}\tilde{T}_\mathrm{eff}=77.7\pm2.0$ eV is known due to pre-outburst
observations in 2003 and 2009 \citep{Ootes_19}. This makes it one of the
hottest NSs in SXTs for their mean accretion rates. Another remarkable
peculiarity of this NS is its spin period of 90 ms
\citep{Papitto_11}, 
which corresponds to order of magnitude slower rotation, than in other
SXTs with known NS spin periods.

The hot quasi-equilibrium state of this NS prompts that it does not
experience the rapid cooling via the direct Urca processes. For the EoS
model BSk24, which we use in the core of the NS, these processes work
for stellar masses above $1.595\,M_\odot$ \citep{Pearson_18}. In the
following we adopt a slightly smaller mass than in Section~\ref{sec:MXB},
$M=1.59\,M_\odot$, which is below this threshold. Otherwise the
simulations of the thermal evolution are performed in
the same way as in Section~\ref{sec:MXB}. 

\begin{figure}[t]
\centering
\includegraphics[width=\columnwidth]{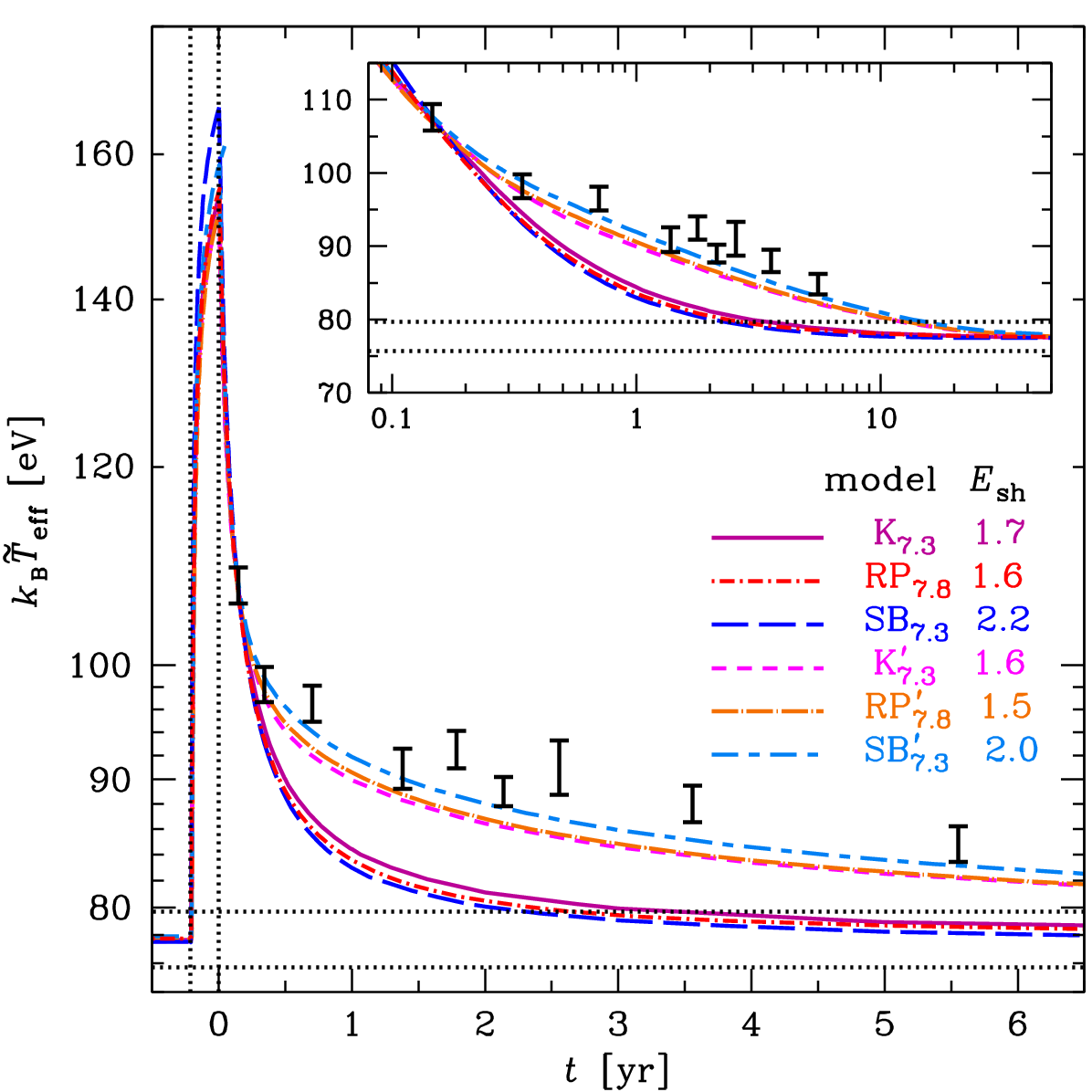}
\caption{Simulated light curves for the NS in  IGR J17480$-$2446
compared with observations of its crust cooling. The models K$_{7.3}$,
RP$_{7.8}$, and SB$_{7.3}$ show the SGC models with
$P_\mathrm{oi}=P_\mathrm{oi}^\mathrm{(0)}$, analogous to  such models in
Fig.~\ref{fig:MXBa}. The models K$_{7.3}'$, RP$_{7.8}'$, and SB$_{7.3}'$
are basically the same, but a layer with a suppressed thermal
conductivity is introduced at $3\times10^{10}$ g cm$^{-3} < \rho <
3\times10^{11}$ g cm$^{-3}$ (see text). The inset shows the same
dependences  and data at another scale for the crust cooling stage only.
The vertical doted lines mark the start and end of the outburst. The
errorbars show the effective temperature observed after the outburst and
the horizontal dotted lines mark the bounds on the quasi-equilibrium
effective temperature observed before the outburst, according to
\citet{Ootes_19} (at the $1\sigma$ confidence level).
}
\label{fig:IGR}
\end{figure}

The results are shown in Fig.~\ref{fig:IGR}. The first three lines (without a
prime in the legend) are for the SGC models K, RP, and SB with
$P_\mathrm{oi}=P_\mathrm{oi}^\mathrm{(0)}$. While the first
observational result (at $t=0.15$ yr) is matched by adjusting
$E_\mathrm{sh}$, these crust cooling curves strongly underestimate
 $\tilde{T}_\mathrm{eff}$ for the later observations.

The slow post-outburst cooling is a known problem for this source.
It was first noted by \citet{Ootes_19}, who analyzed the crust cooling
of IGR J17480$-$2446 using a traditional \citep{HaenselZdunik08} model
of the accreted crust. These authors managed to achieve an agreement
between the theory and observations by introducing a layer with an
unusually low thermal conductivity at densities $10^{11}$ g cm$^{-3} <
\rho < 10^{12}$ g cm$^{-3}$. The suppression of the conductivity was
implemented by using a large impurity parameter $Q_\mathrm{imp}$ in this
layer. They obtained an acceptable agreement between the theory and
observations when using $Q_\mathrm{imp}\gtrsim500$, the best-fit values
being around $Q_\mathrm{imp}\sim1500$.

The suppression of conductivity in the crust may arise due to the lack
of long order in positions of the ions, as in an amorphous solid (a
glass). The phonon spectrum is strongly disordered in this case. We
therefore set a lower limit to the conductivity by treating the ions as
if they were liquefied \citep[cf.][]{Brown00}. This can be roughly
modeled by setting $Q_\mathrm{imp}=\langle Z\rangle^2$.
From physical point of view, it corresponds to the treatment of all ions
as independent scatterers, and within the kinetic theory it is equivalent to
the assumption that the structure factor $S\equiv 1$. The smallest
discrepancy between the observations of IGR J17480$-$2446 and the crust
cooling curves based on the SGC models is obtained, if such an amorphous
layer is located between the mass densities $\sim3\times10^{10}$ g
cm$^{-3}$ and $\sim3\times10^{11}$ g cm$^{-3}$, that is near the bottom
of the outer crust. The light curves, simulated with such an amorphous
layer, are marked in the legend of Fig.~\ref{fig:IGR} by prime
(K$^\prime$, RP$^\prime$, SB$^\prime$). We note in passing that models
assuming the entire outer crust in the amorphous state, as well as those
with an amorphous inner crust (as proposed by \citealt{Jones99,Jones01})
did not provide us with an acceptable agreement between our crust cooling
simulations and the observations.

\section{Conclusions}
\label{sec:concl}

We employed the new set of NS accreted crust models SGC
\citep{ShchechilinGC21,ShchechilinGC22,ShchechilinGC23}, obtained for
realistic nuclear mixtures under the nHD condition, to simulate heating
and cooling of NSs during and after outbursts of SXTs MXB 1659$-$29 and
IGR J17480$-$2446. Unlike the past analyses, which treated the impurity
parameter of the crust $Q_\mathrm{imp}$ as free (sometimes
density-dependent) fitting parameter and where the composition of the
outer heat-blanketing NS envelopes arbitrarily varied, we consistently
adopted these ingredients of the cooling  theory directly from the
underlying SGC models, thus reducing the number of degrees of freedom in
the fitting.

The observed cooling episodes for MXB 1659$-$29 \citep{Parikh_19} were
analyzed consistently, with NS crust parameters for the last cooling
episode tied to those for the previous one.  We found that the SB models
can generate the crust cooling curves that are compatible with the
results of observations of this SXT. As well as in the previous studies,
this agreement between the theory and observations requires an
adjustable heating source at a shallow depth (the shallow heating), in
addition to the deep crustal heating predicted by the theory.

A previous analysis of the observed crust cooling of
the NS in IGR J17480$-$2446 \citep{Ootes_19} revealed that the theoretical cooling
curves could not match observations without introducing a restricted
layer with strongly suppressed thermal conductivity near the outer-inner
crust transition. We eliminate the freedom in the conductivity
suppression by assuming that it arises from a structural disorder,
rather than from the difference in the nuclear composition. Thus we continue to
use the SGC models of the crust composition, but assume that some layer 
of the crust is amorphous and 
has a low conductivity, which is estimated assuming that ions are
independent scatterers ($Q_\mathrm{imp}=\langle Z\rangle^2$).
 In our modeling,
the best-fit position of this layer is estimated to be near the
bottom of the outer crust.

Thus the SGC models are similar to the traditional models of the
accreted NS crusts in their capability to explain the thermal evolution
of NSs in SXTs. Both the traditional and SGC models can serve as the
base for constructing the heating and cooling curves of such NSs
compatible with observations, but only with inclusion of additional
ingredients not supplied by the current theory, such as the shallow
heating and reduction of thermal conductivity.

\section*{Funding}

This research is partly supported (A.Y.P., A.I.C., M.E.G.) by the
Russian Science Foundation Grant No.~22-12-00048.
The work of N.N.S. was financially supported by the FWO (Belgium) 
and the Fonds de la Recherche Scientifique (Belgium) under 
the Excellence of Science (EOS) programme (project No. 40007501).

\section*{CRediT authorship contribution statement}

\textbf{A.Y.~Potekhin:} Conceptualization, Methodology, Software --
simulation of NS structure and thermal evolution, Formal analysis,
Investigation, Visualization, Writing -- original draft, Writing --
review \& editing.
\textbf{A.I.~Chugunov:} Conceptualization, Investigation, Writing --
review \& editing, Project administration.
\textbf{N.N.~Shchechilin:} Software -- calculation of accreted crust
composition and heating rates, Formal analysis, Investigation, Writing
-- review \& editing.
\textbf{M.E.~Gusakov:} Conceptualization, Software -- calculation of
accreted crust composition and heating rates, Writing -- review \&
editing.

\section*{Declaration of competing interest}

The authors declare that they have no known competing financial
interests or personal relationships that could have appeared to influence
the work reported in this paper.

\section*{Data availability}

The tables of the NS crust parameters $\langle Z\rangle$,  $\langle
A\rangle$, $Q_\mathrm{imp}$, and heat release per accreted baryon
$E_\mathrm{h}$ for the SGC models, which are employed in our
calculations, are available at the URL
\href{http://www.ioffe.ru/astro/NSG/accrust/}{http://www.ioffe.ru/astro/NSG/accrust/}.
The other data pertinent to this research will be made available on
request.

\appendix
\setcounter{figure}{0}

\section{Analytical fits for the accreted inner crust}
\label{sec:fits}

We have constructed analytical fitting formulas for evaluation of the
pressure $P$, density $\rho$, and fraction of free neutrons
$Y_\mathrm{nf}$ in the inner crust, according to the SGC models. The
original numerical data for each of the three models (K, RP, and SB) and
each value of $P_\mathrm{oi}$ ($6.4\times10^{29}\mbox{~dyn cm}^{-2} \leq
P_\mathrm{oi} \leq 10^{30}\mbox{~dyn cm}^{-2}$ with the step $\Delta
P_\mathrm{oi} = 10^{28}$ dyn cm$^{-2}$) cover the density range
$\rho_\mathrm{oi+} \leq \rho \leq
\rho_\mathrm{max}=2\times10^{12}\mbox{~g cm}^{-3}$, where
$\rho_\mathrm{oi+}$ is the density at the upper edge of the inner crust
(at $P=P_\mathrm{oi}$). In this density range, pressure logarithm is
fitted by the formula
\begin{equation}
   \log(P/P_\mathrm{oi}) = a_1 x + b\, x^3,
\label{fitP}
\end{equation}
where
\begin{equation}
   x = \log\,(\rho/\rho_\mathrm{oi+}),
\quad
   b = a_2 + a_3\log\,P_\mathrm{oi29},
\end{equation}
$P_\mathrm{oi29}\equiv
P_\mathrm{oi}/(10^{29}\mbox{~dyn cm}^{-2})$, and the fit parameters
$a_i$ are given in Table~\ref{tab:fitP}. The fit reproduces the
calculated $P$ values
with a typical accuracy $\sim1\%$ and a maximum discrepancy of 2\% for
all three model families and all 37 values of $P_\mathrm{oi}$.

Equation (\ref{fitP}) has the form of a depressed cubic equation, which
allows an analytical inversion, providing a fit of $\rho$ as a
function of $P$. The maximum error of this inverse fit lies within 3.5\%.

\begin{table}[t]
\centering
\caption{Fit parameters in Eq.~(\ref{fitP})}
\label{tab:fitP}
\begin{tabular}{l c c c}
\hline
Ashes model& $a_1$   & $a_2$ & $a_3$ \\ 
\hline
Kepler     & 0.494 & $-0.278$ & 0.760 \\
Extreme rp & 0.507 & $-0.376$ & 0.861 \\
Superburst & 0.488 & ~~~0.021 & 0.433 \\
\hline
\end{tabular}
\end{table}

The boundary density $\rho_\mathrm{oi+}$ in Eq.~(\ref{fitP}) is the
upper end of the density jump at the transition from the outer crust to the
inner crust.  The lower end of this jump, $\rho_\mathrm{oi-}$, can
be considered as the bottom density of the outer crust. Both these
densities are reproduced as functions of $P_\mathrm{oi}$ by the
piecewise power law
\begin{equation}
   \rho_\mathrm{oi\pm} = c_\pm P_\mathrm{oi29}^{3/4},
\label{oi}
\end{equation}
where the factors $c_\pm$ are constant in certain intervals of
$P_\mathrm{oi}$. The values of $c_\pm$ in Table~\ref{tab:oi} provide
the accuracy of $\rho_\mathrm{oi\pm}$ within 0.15\% at most (typically
$\sim\mbox{a few}\times0.01\%$).

\begin{table}[t]
\centering
\caption{Values of factors $c_\pm$ in Eq.~(\ref{oi}), in units of
$10^{11}$ g cm$^{-3}$.}
\label{tab:oi}
\begin{tabular}{c c | c c}
\noalign{\smallskip}
\hline
$P_\mathrm{oi29}$ & $c_-$ & $P_\mathrm{oi29}$ & $c_+$  \\ 
\hline
\multicolumn{4}{c}{Kepler} \\
6.4 -- 7.8  & 0.8577 & 6.4 -- 8.3  & 0.9057  \\
7.9 -- 8.8  & 0.8627 & 8.4 -- 8.9  & 0.9251  \\
     8.9    & 0.8671 & 9.0 -- 10   & 0.9224  \\
9.0 -- 9.5  & 0.8834 & -- & -- \\
9.6 -- 10   & 0.8865 & -- & -- \\
\hline
\multicolumn{4}{c}{Extreme rp} \\
6.4 -- 6.7  & 0.8658 & 6.4 -- 7.9  & 0.9105 \\
6.8 -- 7.2  & 0.8679 & 8.0 -- 8.3  & 0.9333 \\
7.3 -- 7.5  & 0.8868 & 8.4 -- 9.5  & 0.9365 \\
7.6 -- 8.4  & 0.9008 & 9.6 -- 10   & 0.9396 \\
8.5 -- 9.2  & 0.9038 &  -- & -- \\
9.3 -- 9.6  & 0.9057 &  -- & -- \\
9.7 -- 10   & 0.9075 &  -- & -- \\
\hline
\multicolumn{4}{c}{Superburst} \\
6.4         & 0.8410 & 6.4 -- 6.5  & 0.9024 \\
6.5 -- 7.5  & 0.8630 & 6.6 -- 10   & 0.9112 \\
7.6 -- 10   & 0.8669 &  -- & -- \\
\hline
\end{tabular}
\end{table}

\begin{figure}[t]
\centering
\includegraphics[width=\columnwidth]{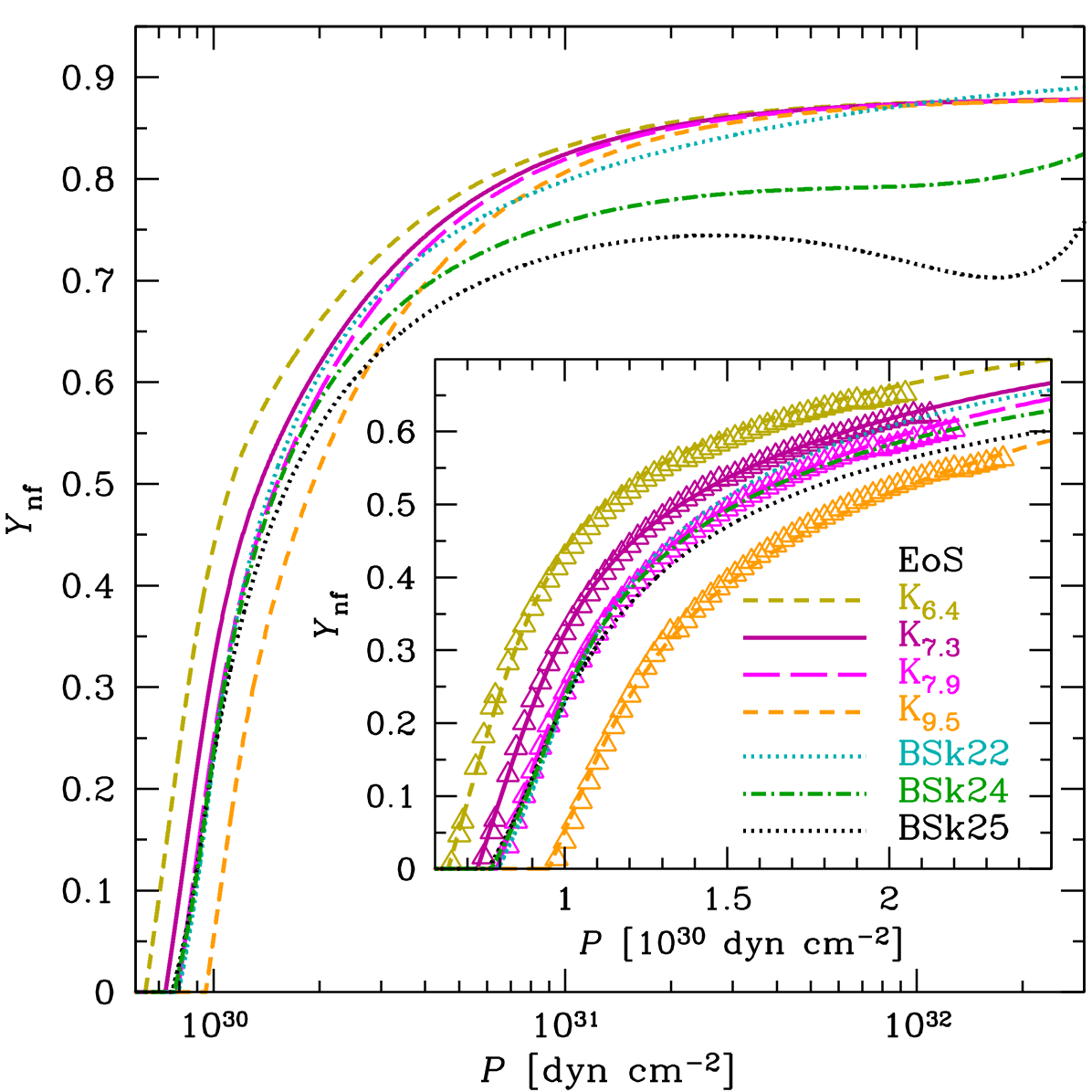}
\caption{Dependence of the fraction of free nucleons $Y_\mathrm{nf}$  in
the inner crust of a NS on pressure. The inset presents a zoom
to the upper part of  the inner crust. The lines correspond to
analytical approximations, according to the legend: four lines for the
SGC accreted crust with $P_\mathrm{oi29} = 6.4$, 7.3, 7.9, and 9.5, and
three ground-state models from the Brussels--Montreal family -- BSk22,
BSk24, and BSk25. The open triangles in the inset show the numerical SGC
results for the K model.
}
\label{fig:Ynf}
\end{figure}

The fraction of free neutrons relative to all nucleons $Y_\mathrm{nf}$
can be fitted as a function of pressure $P$
(at $P > P_\mathrm{oi}$, for  all SGC models) as
\begin{equation}
   Y_\mathrm{nf} = \frac{y}{1+ay^3/(1+0.88ay^2)},
\label{Ynf}
\end{equation}
where
\begin{equation}
   y = \frac{P}{P_\mathrm{oi}}-1,
\quad
   a=0.011 P_\mathrm{oi29}^3.
\end{equation}
The absolute maximum difference of this approximation from the numerical
data for all the $37\times3$ considered versions of the SGC models
is $\approx0.02$. The approximation (\ref{Ynf}) is shown in
Fig.~\ref{fig:Ynf}. Here for illustration we show the K model with
$P_\mathrm{oi}\approx P_\mathrm{oi}^{(0)}$
($P_\mathrm{oi29}=7.3$), with $P_\mathrm{oi}\approx P_\mathrm{oi}^\mathrm{(cat)}$
($P_\mathrm{oi29}=7.9$), and with substantially lower and
higher values of $P_\mathrm{oi}$ ($P_\mathrm{oi29}=6.4$ and 9.5). The
inset shows the data (triangles) and the fit (lines) at
$\rho_\mathrm{oi} \lesssim \rho \lesssim \rho_\mathrm{max}$, while the
main window shows the fit and its extrapolation to the crust
bottom. For comparison, Fig.~\ref{fig:Ynf} also shows $Y_\mathrm{nf}$ in
three representative models of cold catalyzed crust: BSk22, BSk24, and
BSk25 \citep{Pearson_18}. The comparison shows that the extrapolation of
the fit is not unreasonable.


\newcommand{\arnps}{Annu.\ Rev.\ Nucl.\ Part.\ Sci.}
\newcommand{\al}{Astron.\ Lett.}
\newcommand{\aap}{Astron.\ Astrophys.}
\newcommand{\apj}{Astrophys.\ J.}
\newcommand{\apjl}{Astrophys.\ J.}
\newcommand{\apjs}{Astrophys.\ J.\ Suppl.\ Ser.}
\newcommand{\epja}{Eur.\ Phys.\ J. A}
\newcommand{\iaucirc}{IAU Circ.}
\newcommand{\jaa}{J.\ Astron.\ Astrophys.}
\newcommand{\nat}{Nature}
\newcommand{\mnras}{Mon.\ Not.\ R.\ Astron.\ Soc.}
\newcommand{\npa}{Nucl.\ Phys. A}
\newcommand{\physrep}{Phys.\ Rep.}
\newcommand{\prc}{Phys.\ Rev. C}
\newcommand{\prd}{Phys.\ Rev. D}
\newcommand{\prl}{Phys.\ Rev.\ Lett.}
\newcommand{\ptp}{Prog.\ Theor.\ Phys.}
\newcommand{\ppnp}{Prog.\ Part.\ Nucl.\ Phys.}
\newcommand{\rmp}{Rev.\ Mod.\ Phys.}
\newcommand{\ssr}{Space Sci.\ Rev.}

\end{document}